\documentclass[prl,preprint,onecolumn,showpacs,preprintnumbers,amsmath,amssymb,floatfix]{revtex4}
\usepackage[german,english]{babel}
\usepackage{bm,amsmath,amsfonts}
\usepackage{amssymb,siunitx}
\usepackage{graphicx}
\usepackage{xcolor}

\def\beq{\begin{equation}}
\def\eeq{\end{equation}}
\def\beqn{\begin{eqnarray}}
\def\eeqn{\end{eqnarray}}

\def\brho {\mbox{\boldmath $\rho$}}
\def\r {\vec{\bf r}}
\def\r {{\bf r}}
\def\br {{\bf r}}
\def\bR {{\bf R}}

\def\beq{\begin{equation}}
\def\eeq{\end{equation}}
\def\beqn{\begin{eqnarray}}
\def\eeqn{\end{eqnarray}}

\def\r {{\bf r}}

\def\C {{\bf C}}

\def\r {\vec{\bf r}}
\def\r {{\bf r}}
\def\br {{\bf r}}
\def\bR {{\bf R}}

\def\brho {\mbox{\boldmath $\rho$}}

\begin{document}

\title{Quantum systems of ultra-cold bosons with customized inter-particle interactions}

\author{Alexej I. Streltsov}
\affiliation{Theoretische Chemie, Physikalisch-Chemisches Institut, Universit\"at Heidelberg,\\
Im Neuenheimer Feld 229, D-69120 Heidelberg, Germany}

\begin{abstract}
Recent progress in cooling and trapping of  polarized clouds of chromium $^{52}Cr$, 
dysprosium $^{164}Dy$ and erbium $^{168}Er$ opens a road-map to quantum systems
where shapes of inter-particle interactions can be customized.
The main purpose of this work is to get a deeper insight on a role which the overall shape of the inter-particle interaction plays
in a context of trapped ultra-cold bosons. We show that strong inter-particle repulsion inevitably leads
to multi-hump fragmentation of the ground state. The fragmentation phenomenon is universal -- 
it takes place in traps of different dimensionality and 
topologies and for very broad classes of repulsive inter-particle potentials. The physics behind is identified and explained.
\end{abstract}


\maketitle

Nowadays, dilute ultra-cold atomic/molecular clouds are considered as toolboxes to probe static and dynamical
properties of many-particle Hamiltonians \cite{Pitaevskii_review,Leggett_review,COLD_MOL_Rev}.
Consequently, some phenomena which are very difficult or even impossible to study in their natural appearance/environment 
can explicitly be reconstructed and modeled in the ultra-cold atomic systems.
However, until recent times one substantial ingredient was missed - a control on the overall shape of inter-particle interactions.

Recent experiments with ultra-cold polarized clouds of chromium $^{52}Cr$ \cite{Cr_exp_1,Cr_exp_2},
dysprosium $^{164}Dy$ \cite{Dy_exp_1} and erbium $^{168}Er$  \cite{Er_exp_1}
have ultimately shown that the short-range inter-particle interaction 
potential alone cannot describe the observed physics and it should be augmented by an additional long-range term
which usually takes on the form of a dipole-dipole interaction \cite{Dip_review0,Dip_review}.
Another venue to manipulate the  effective inter-particle interaction is to 
admix a small component of an excited Rydberg state to the ground state of ultra-cold atoms \cite{Ry1,Ry2}
via the off-resonant optical coupling, creating thereby so-called ``dressed" Rydberg systems.
These steps towards control and manipulation of the overall shape 
of the inter-particle interaction open a road-map to mimic/simulate static 
and dynamical phenomena and effects appearing in the context of other fields of physics 
where shapes of inter-particle interactions play crucial roles, e.g. in nuclear physics.

The perspective of working with quantum systems where the inter-particle interaction is customized encourages us to 
get a deeper insight on a role it plays.
The role of the sign is evident: if it is negative -- the system is attractive, if it  is positive --  repulsive,
but what roles are playing its range and tails? What physical phenomena or properties 
they are envisioned to impact?
The main phenomenon predicted and intensively studied in the field of ultra-cold atoms
is condensation, manifesting itself in a peaked density 
with a profile similar to the lowest-in-energy eigenstate of the confining potential. 
Condensation also means coherence between weakly-interacting particles ~\cite{Pitaevskii_review,Leggett_review}.
On the other hand, the existence in the ground-state density of multi-hump features 
might indicate on possible (strong) correlations in the system which usually destroy condensation and 
break partially or completely coherence between the particles.

The multi-hump structure of the ground-state density can be caused by applied external potential barriers.
In a double-well for instance, a sufficiently high barrier can split the density into two well-separated sub-clouds.
Such a system is then called two-fold fragmented \cite{frag1} implying that coherence between the sub-fragments is lost.
If more barriers are available, as in optical lattices, the system can be multi-fold fragmented, see in this respect 
the famous Mott-insulating phases  \cite{MI}.
Complimentary, modulations of density profiles and loss of an inter-particle coherence
can be caused by strong repulsive inter-particle interaction.
In one-dimensional systems with contact inter-particle interactions
it originates to the famous fermionization phenomenon \cite{Girardeau_TG,Fermionization_Exp}
for inter-boson interactions of other shapes --  to solid-like states see, 
e.g. Coulomb bosons \cite{1overR_1D_1} and dipolar (screened Coulomb-like) bosons \cite{1overR3_1D}.
In two-dimensions strong repulsion is predicted to be responsible for so-called crystallization \cite{Dip_review0,Dip_review} --
appearances of stable rims in the density profiles and/or its partial factorization into multi-hump structures,
see e.g. Refs.~\cite{PP_dipolar_2D_1,PP_dipolar_2D_2}.

The main goal of the present work is to investigate the microscopic details of how repulsive inter-particle potentials 
create the non-trivial features (humps) in the densities of ultra-cold systems confined in simple barrier-less traps.
We also want to understand which characteristic of a general inter-particle interaction function
favors these density modulations and controls, thereby, the accompanying 
developments of correlations and fragmentation.

Let us consider a generic many-body Hamiltonian of $N$ identical bosons
trapped in an external trap potential $V(\r)$ and interacting via general inter-particle interaction potential:
$
\hat H(\r_1,\ldots,\r_N) \!=\! \sum_{j=1}^N \left[- \frac{1}{2} \nabla^2_{\r_j} \!+\! V(\r_j) \right]
\!+\!\sum_{j < k}^N \lambda_0 W(\r_j\!-\!\r_k). \nonumber
$
Here $\lambda_0$ defines the strength of the interaction and  $W(\br\!-\!\br') \equiv W(\bR)$ its shape.
In this work $\hbar=1$, $m=1$.

To start with we consider several one-dimensional barrier-less traps:
a standard harmonic $V(x)\!=\!0.5x^2$, a non-harmonic $V(x)\!=\!0.5x^6$ and
asymmetric linear $V(x)\!\!=\!\!\left\{-x:x\!<\!0;3x/4\!:\!x\!\ge\!0\right\}$. 
We examine the following inter-particle interaction functions:
exponential $\mathrm{exp}\left[-\frac{1}{2}(|x-x'|/D)^n\right]$,  screened Coulomb $1/\sqrt{(|x-x'|/D)^{2n}+1})$ and
Sech-shaped $\mathrm{sech}\left[(|x-x'|/D)^n\right]$ of half-width $D$ with $n=1$ and their \emph{sharper} analogous with $n=2$. 
In the following we use the shorthand notations $\mathbf{exp}\left[-\bR^n\right]$, $1/\bR^n$ and $\mathbf{sech}\left[\bR^n\right]$.
To solve the respective many-boson Schr{\"o}dinger equation numerically we use 
the recently developed Multi-Configurational Time-Dependent Hartree method for Bosons  (MCTDHB)  \cite{ramp_up,MCTDHB_paper},
see \cite{SM} for more details. 
This method is capable of providing numerically-exact solutions \cite{MCTDHB_HIM}.

\begin{figure}[]
\includegraphics[width=17cm,angle=0]{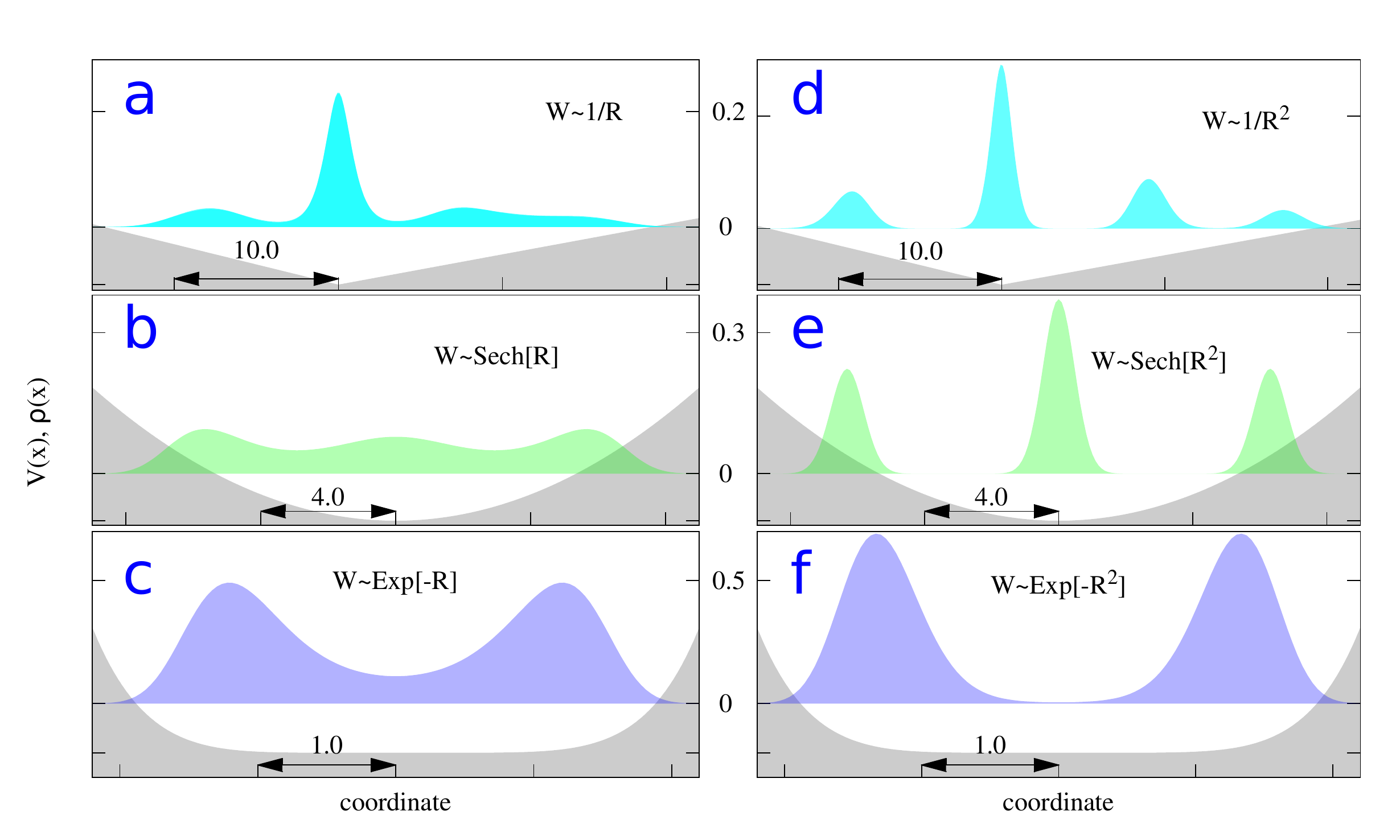}
\caption{(color online). Diversity of fragmentation phenomena.
Strong inter-particle repulsion leads to formation of multi-hump localized, fragmented  structures
irrespective to the shapes of inter-particle $W(\bR)$ and trapping $V(\br)$ potentials used.
Shown are the densities of one-dimensional systems made of $N=108$ bosons
and corresponding trapping potentials which are scaled and shifted for better presentation.
Upper panels: asymmetric linear trap and screened Coulomb inter-boson interactions 
of half-width $D=5$ and strength $\lambda_0=0.3$ with $n=1$ ({\bf a}) and $n=2$ ({\bf d}).
Middle panels: harmonic trap and  Sech-shaped interactions
with $D=4$ and $\lambda_0=1.0$,  $n=1$ ({\bf b}), $n=2$ ({\bf e}).
Lower panels: non-harmonic trap and exponential inter-particle interactions  
with $D=3$ and $\lambda_0=1.5$, $n=1$ ({\bf c}), $n=2$ ({\bf f}).
Inter-particle interaction potentials with \emph{sharper} edges ($n=2$) enhance fragmentation, see text for details.
All quantities shown are dimensionless.}
    \label{fig1}
\end{figure}

The left panels of Fig.~\ref{fig1} plot the ground-state densities obtained for 
one-dimensional systems with 
$1/\bR$, $\mathbf{sech}\left[\bR\right]$, $\mathbf{exp}\left[-\bR\right]$ inter-particle interaction functions of different widths (ranges) 
confined in different traps. The main fascinating observation is that 
irrespective to the shapes of inter-particle and trapping potentials used
a strong enough inter-particle repulsion leads to formation of the multi-hump localized structures  and, therefore,
indicates on possible correlations/fragmentation in the systems.

\begin{figure}[]
\includegraphics[width=17cm,angle=0]{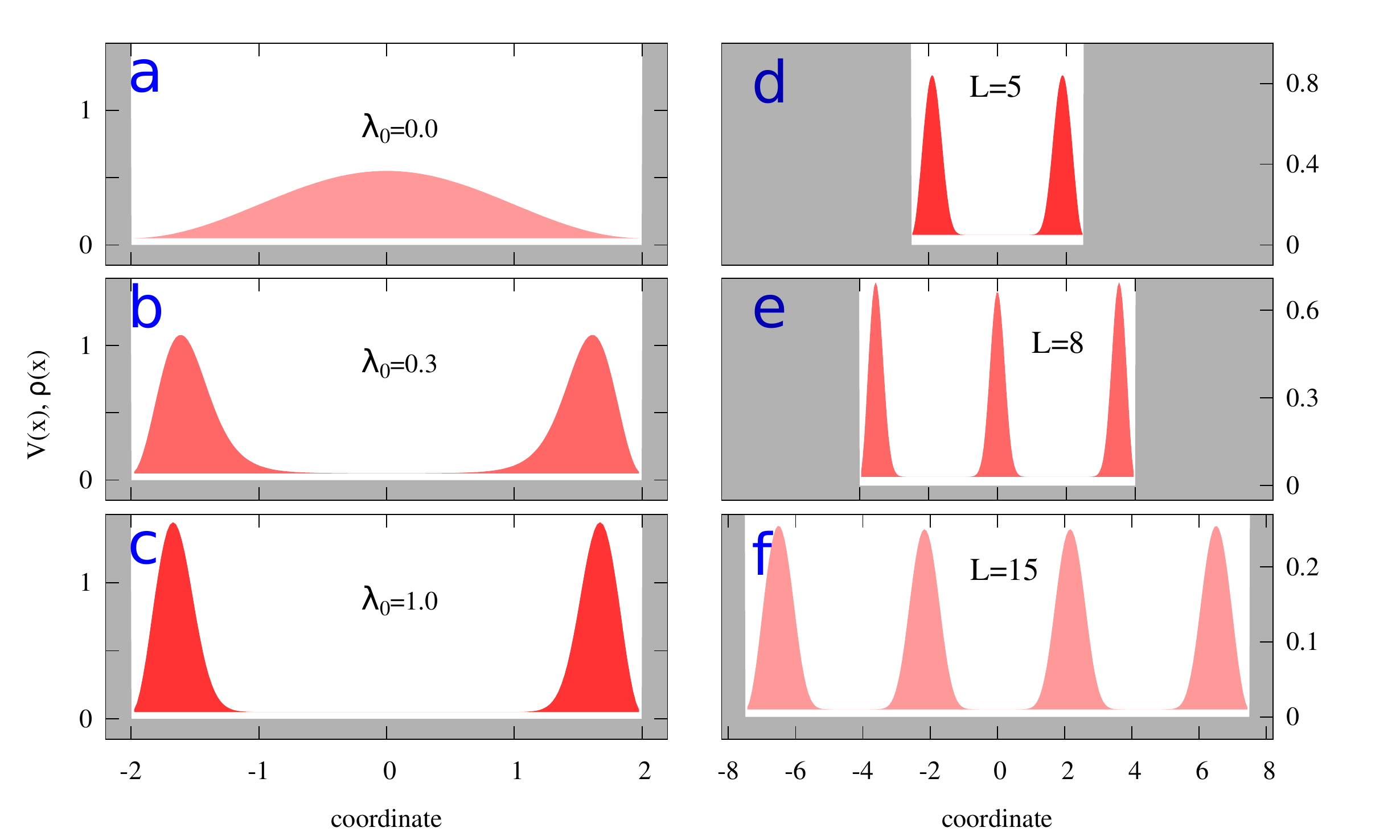}
\caption{(color online). Physics of fragmentation in bosonic systems confined in box-shaped traps of length $L$
and interacting via inter-particle interaction potential of rectangular shape with half-width $D=3$.
The left panels show that for strong repulsion $\lambda_0=0.3,1.0$ ({\bf b,c})  
it costs energy to keep $N=108$ bosons in the middle of the trap ($L=4$) and to minimize the energy
the system fragments into two well-separated sub-clouds.
The right panels show how to control the fragmentation in strongly repulsive systems ($\lambda_0=1.0$)
by varying the length $L$ of the box.
For $L=8$ the ground state is three-fold fragmented ({\bf e}) and for $L=15$ -- four-fold ({\bf f}). 
Shown are the one-particle densities and trap potentials. 
All quantities shown are dimensionless.}
    \label{fig2}
\end{figure}

To shed more light on what characteristic parameters of the inter-particle interaction function
drive and control the number of humps and correlations/fragmentation
we have computed the ground-states' properties of the same systems as above, 
but with \emph{sharper} inter-particle interactions  $1/\bR^2$, $\mathbf{sech}\left[\bR^2\right]$, $\mathbf{exp}\left[-\bR^2\right]$.
The results presented in the right panels of Fig.~\ref{fig1} show 
that for sharper versions of the inter-particle interactions
the multi-hump structures are retained and become even more pronounced.
The many-body analysis of these multi-hump solutions
reveals that they are not condensed but multi-fold fragmented,
see \cite{SM} for more details.
In the asymmetric linear trap the four-hump density of $N=108$ bosons depicted in Fig.~\ref{fig1}d
is formed by four contributing natural sub-fragments $\rho(x)\!=\!\sum^4_{i=1}n_i |\phi^{NO}_i(x)|^2$.
The first fragment with $n_1=52$ bosons is localized at the trap minimum and forms the most intense hump.
The left-most hump is formed by $n_3=20$ bosons residing in a well-localized third natural orbital.
Two other humps at the right are formed by $n_2\approx27.6$ and $n_4\approx8.4$ bosons
residing in the second and fourth natural orbitals which are slightly delocalized.
In the harmonic trap and Sech-shaped interaction, see Fig.~\ref{fig1}e, the three symmetric fragments carry 30, 48, and 30 bosons, respectively.
In the non-harmonic trap a Gaussian interaction results in 
a two-fold fragmented ground state with 97.2 and 10.8 bosons per fragment, see Fig.~\ref{fig1}f.
Here it is worthwhile to stress that for weak repulsions the regime of normal condensation is, of course, recovered
irrespective to the particular form of interactions, see \cite{SM} for more details.

The above observed diversity of the fragmentation is caused by the interplay between the inter-particle and trapping potentials.
To distinguish and isolate effects originating 
from the range/width of the inter-particle interaction, particular shape of its tails 
and strength of the inter-particle repulsion let us consider a box-shaped trap and inter-particle interaction potential of a rectangular shape
$W\left(x\!-\!x'\right)\!=\!\left\{1: |x\!-\!x'| \leq D; \, 0: \mathit{otherwise} \right\}$.
In this system an increase of the strength of the inter-particle repulsion does not change 
neither the effective length of the trap nor the range of the inter-particle interaction potential.

The left panels of Fig.~\ref{fig2} show how the ground-state densities of the system made of $N=108$ bosons
with rectangular inter-particle interaction of half-width $D=3$, trapped in a box-shaped trap of length $L=4$, 
response when the repulsion strength is increased $\lambda_0=0,0.3,0.6$.
We see that in the non-interacting case the density is broad and has a maximum at the trap center.
The strong inter-particle repulsion, in contrast, leads to localization of the density at the edges of the box.
Basic ``electrostatic'' arguments can explain this behavior.
For strong repulsion it costs energy to keep the bosons in the middle of the trap,
so, they repeal and push each other away -- the cloud starts to form a minimum in the center of the trap.
If we increase the repulsion further -- the total energy is increased, but the system can not 
expand further apart due to the box-shaped topology of the trap. 
To minimize the energy the density is split into two well-separated fragments.

The right panels of Fig.~\ref{fig2} present a complimentary study where
we keep the inter-particle interaction $\lambda_0=1.0$, $D=3$ fixed and increase the box size $L=5,8,15$
providing, thereby, more room for the bosons.
We see that starting from some critical length of the trap (box size), to minimize the repulsion
the system of $N=108$ bosons is split into three sub-clouds with 36 bosons per fragment.
To split the system into four fragments with 27 bosons per sub-cloud
one has to increase the trap's length further on.
So, for strong enough inter-particle repulsions 
ground state fragmentation is an inevitable property of trapped bosonic systems.
The interplay between the width of the finite-range part of the inter-particle interaction function and 
the length of the trap defines the particular fragmentation scenario.

\begin{figure}[]
\includegraphics[width=17cm,angle=0]{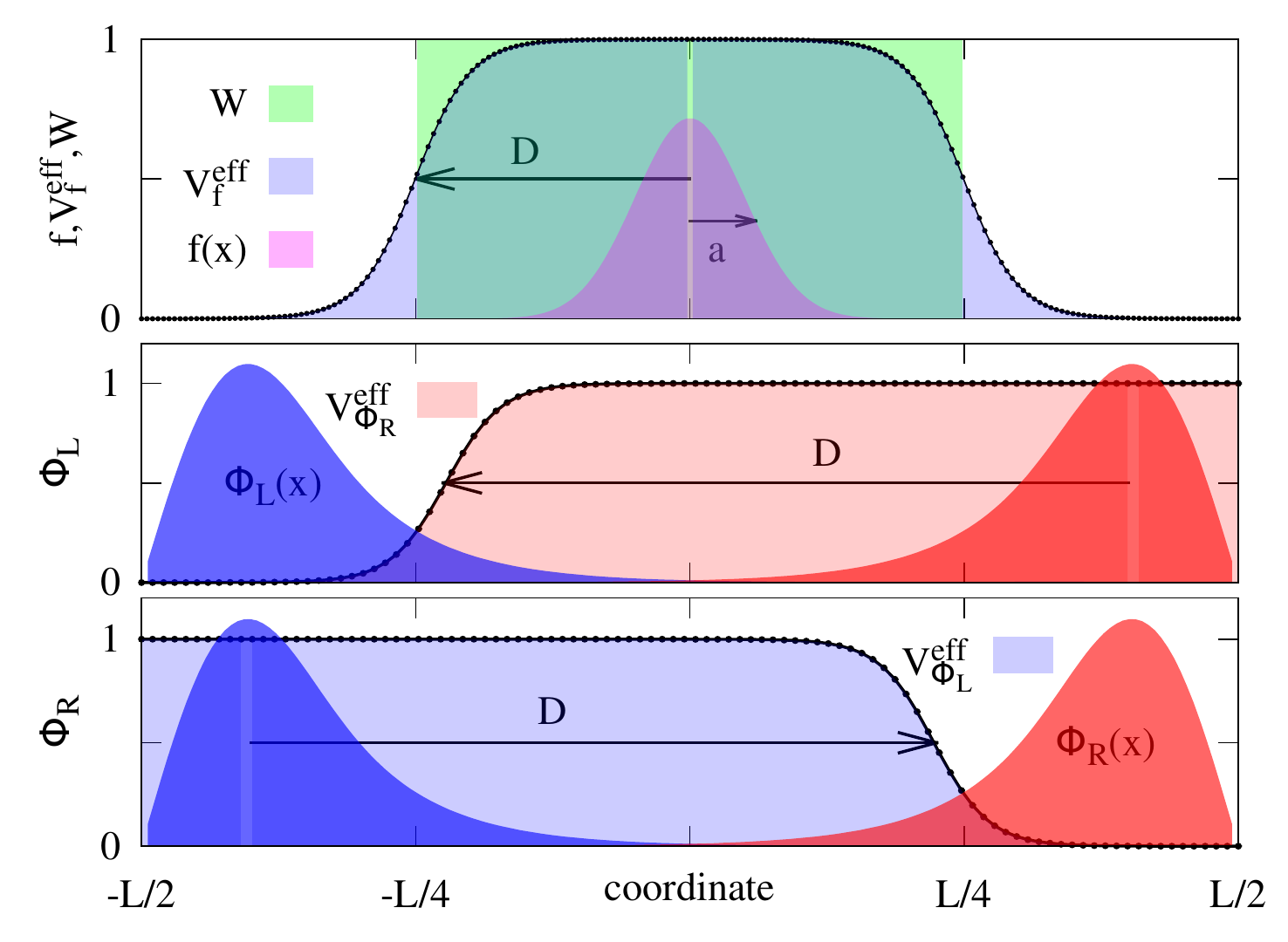}
\caption{(color online). Explanation on how finite-range inter-particle interaction potentials induce effective barriers and confine fragmented states.
The upper panel plots by a solid line with circles an effective barrier $V^{eff}_f(x)\!=\!\int |f(x')|^2 W(x\!-\!x') dx'$
produced by a rectangular inter-particle interaction potential $W(x\!-\!x')$ of half-width $D$ acting
on a Gaussian-shaped function $f(x)$ of half-width $a$. 
Lower panel: a rectangular inter-particle potential acting on the sub-cloud  $\phi_L$ produces
an effective barrier $V^{eff}_{\phi_L}(x)\!=\!\int |\phi_L(x')|^2 W(x\!-\!x') dx'$.
The superposition of this interaction-induced effective barrier and 
the external trap (here it is a square box of length $L$)
results in a well where the $\phi_R$ sub-cloud is confined.
Middle panel: the action of the same finite-range $W(x\!-\!x')$ on $\phi_R$ produces
an effective well capable of confining the left sub-cloud $\phi_L$.
All quantities shown are dimensionless.}
    \label{fig3}
\end{figure}

To get a deeper insight into the physics behind the fragmentation
let us rely on an idealized picture of a two-fold fragmented state.
The total density of such a state is formed by a sum of two isolated and independent sub-clouds (fragments).
Its many-body wave-function is then a Fock state $|n_1,n_2\rangle$ represented by a single symmetrized permanent 
$ \Psi(\br_1,\ldots,\br_N)=\hat{\cal S}\phi_L(\br_1)\cdots\phi_L(\br_{n_L})\phi_R(\br_{n_L+1})\cdots\phi_R(\br_{n_L+n_R})$
with $n_L$ bosons residing in the left  $\phi_L$ and $n_R$ in the right  $\phi_R$ fragment, respectively.
The optimal shapes of the fragments are determined self-consistently by solving  
the multi-orbital best mean-field (BMF) equations \cite{BMF,BMF_pla_2D_3D} which
we rewrite as (see \cite{SM} for details):
$$
\begin{array}{ll}
\!\!\left[ \hat h\!+\!\lambda_0(n_L\!-\!1) V^{eff}_{\phi_L}(\br)\!+\!\lambda_0n_RV^{eff}_{\phi_R}(\br) \right]\!\phi_L\!\!=\!\!
\mu_{11}\phi_L\!+\!\mu_{12}\phi_R   \nonumber  \\ 
\!\!\left[\hat h\!+\!\lambda_0(n_R\!-\!1) V^{eff}_{\phi_R}(\br)\!+\!\lambda_0n_LV^{eff}_{\phi_L}(\br) \right]\!\phi_R\!\!=\!\!
\mu_{21}\phi_L\!+\!\mu_{22}\phi_R. 
\end{array}
$$
Here $\hat{h}\!=\!-\frac12\nabla^2_{\br}\!+\!V(\br)$ is a single-particle Hamiltonian. The $V^{eff}_{\phi_i}(\r)\!=\!\int |\phi_i(\r')|^2 W(\r\!-\!\r') d\r'$ 
terms play the roles of effective \emph{self-consistent} potentials --
their profiles depend on a given shape of inter-particle function  $W(\r\!-\!\r')$  and on the left/right densities $|\phi_i(\r')|^2$, $i\!=\!L,R$
of the involved sub-clouds.

The upper panel of  Fig.~\ref{fig3} exemplifies  schematically how 
effective barrier $V^{eff}_{f}\!=\!\int |f(x')|^2 W(x\!-\!x') dx'$ is produced by
a rectangular inter-particle interaction potential $W(x\!-\!x')$ of range $D$ 
acting on a Gaussian-shaped cloud $f(x)$ of half-width $a$.
This effective barrier depicted by a solid line with circles is centered at the cloud origin.
The profile of this barrier is flat over the extent of the cloud $f(x)$ and drops like $\mathbf{erf}$-function at the $\pm D$ edges.

Two other panels of Fig.~\ref{fig3} show how a rectangular inter-particle interaction of half-width $D$  induces 
the effective potentials in the above studied one-dimensional bosonic system trapped in a box-shaped trap.
The lower panel of Fig.~\ref{fig3} depicts by a solid line with circles
the interaction-induced effective potential $V^{eff}_{\phi_L}(x)\!=\!\int |\phi_L(x')|^2 W(x\!-\!x') dx'$. 
It is constant over the extent of the ${\phi_L}(x)$ fragment and, therefore,
in the above BMF equations
it just shifts the left-localized sub-cloud upwards energetically without changing its shape:
$V^{eff}_{\phi_L}(x){\phi_L}(x)\!\approx\!\int |\phi_L(x')|^2 dx' \phi_L(x)\!=\!\phi_L(x)$.
The action of  $V^{eff}_{\phi_L}(x)$ on the right fragment ${\phi_R}(x)$, in contrast, is dramatic -- 
it induces the effective barrier. A superposition of this effective barrier and the external trap (box)
is capable of confining the right sub-cloud ${\phi_R}(x)$.
Similarly, the action of $W(x\!-\!x')$ on the right sub-cloud creates an effective barrier  $V^{eff}_{\phi_R}(x)$ confining the left fragment,
see middle panel of Fig.~\ref{fig3}.

We have arrived here at a microscopic, self-consistent picture of {\it self-induced} fragmentation which
requires as a prerequisite localized sub-clouds, finite-range inter-particle repulsion and a trap of a finite length.
The action of a finite-range repulsive inter-particle potential on a cloud produces an effective barrier.
The profile of this barrier depends on the density of the cloud, number of the particles in it and on the shape of $W(\br\!-\!\br')$. 
When several localized clouds/fragments are present each of them creates its own effective potential
seen by the other sub-clouds as an effective barrier. The superposition of these self-induced barriers and external trap results 
in a multi-well potential confining the fragmented system as a whole object.

The multi-hump structure of the density is driven by the fragmentation phenomenon 
which has been observed experimentally \cite{Kasevich0,JoergNatPhys06} and 
is well-understood theoretically \cite{frag1,frag2,frag3,frag4}.
From this perspective, variations of shapes of $W(\r\!-\!\r')$ can modify the induced effective potentials
and, thereby, the fragmentation in the following manner:
a) by decreasing range of the inter-particle repulsion we increase the overlap between the fragments which reduces the fragmentation
and leads to the development of coherence between them \cite{build_up_coherence};
b) by increasing strength of the inter-particle repulsion and keeping its range fixed 
we increase the heights of the induced barriers which isolates the fragments and enhances fragmentation;
c) by flattening the tails of the inter-particle interaction function one stimulates the overlap between 
the sub-clouds which melts the humps and blurs the fragmentation. 
Comparison of the left and right panels of Fig.~\ref{fig1} confirms this analysis. 

A simple geometrical relation between the effective length of the trap $L$ and
width/range $D$ of the finite-range part of the inter-particle interaction function
defines the number $M$ of available humps(fragments).
For $L\!\!\!<\!\!\!D$ we have a system of non-interacting particles because irrespective to the shape and size of the trapped cloud,
the created effective barrier is geometrically broader than the trap (box) itself.
When $L\!>\!D$ there is enough room for two fragments to be trapped by the self-induced effective barriers,
to accommodate $M$ fragments the length of the trap should be  $L\!>\!(M-1)D$, see e.g. Fig.~\ref{fig2}.

\begin{figure}[]
\includegraphics[width=8cm,angle=0]{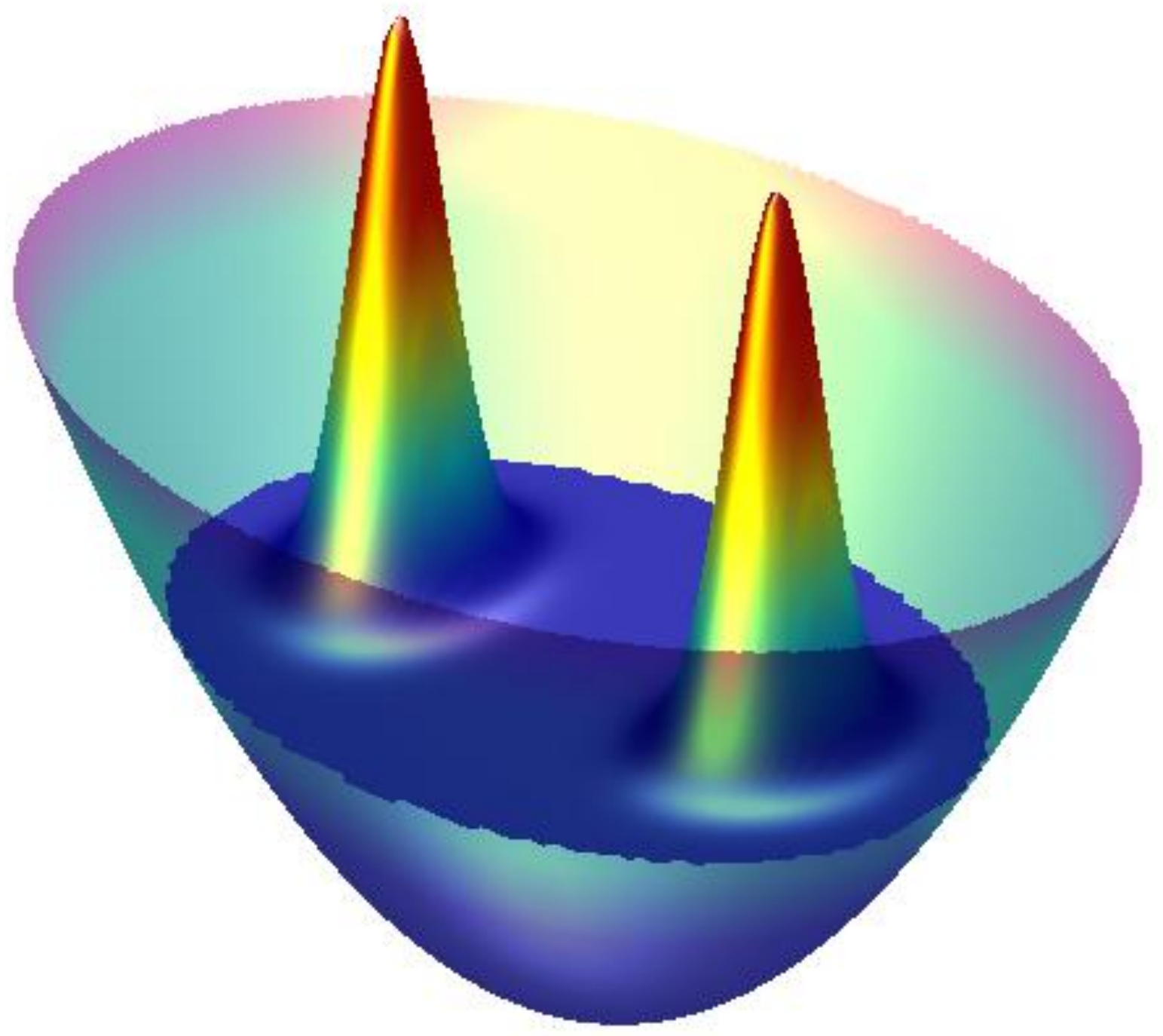}
\includegraphics[width=8cm,angle=0]{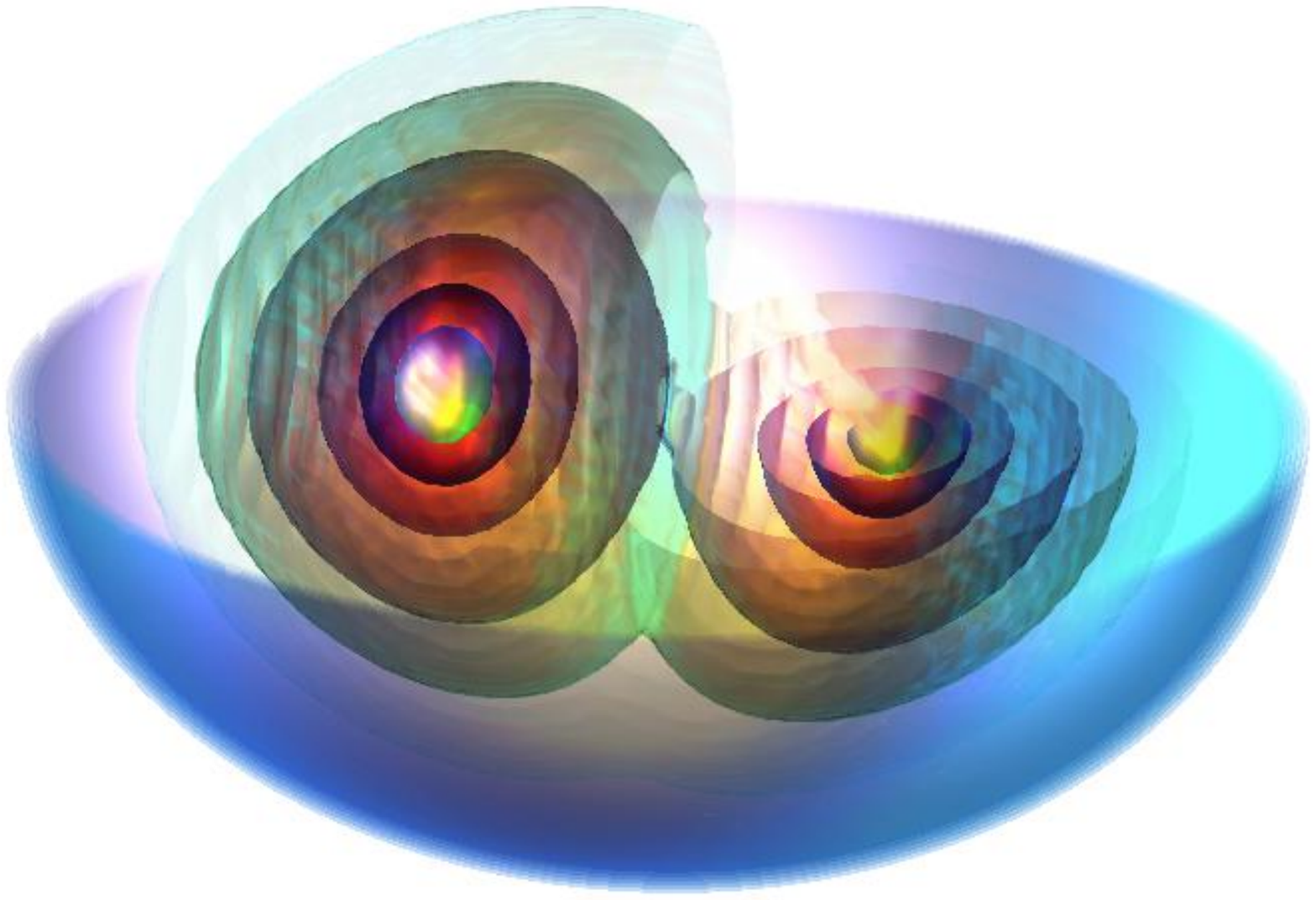}
\caption{(color online). Universality of self-induced fragmentation mediated by finite-range inter-particle interaction potentials.
Left panel: two-fold fragmented ground state in harmonic 2D trap slightly elongated in one direction.
Right panel: two-fold fragmented ground state in a slightly elongated 3D harmonic trap.  
To visualize the 3D functions we plot several isosurfaces of the density and an equipotential cut of the trap. 
All quantities shown are dimensionless.}
    \label{fig4}
\end{figure}

The obtained geometrical picture of 
localized fragments and self-induced effective potentials
is universal and can be applied to systems of bosons, fermions and distinguishable particles also 
at higher dimensions. In the left and right panels of Fig.~\ref{fig4} 
we plot illustrative numerical examples of 
almost ideal self-induced two-fold fragmented ground states of 
bosonic systems with $N=100$ particles
interacting via sharp repulsive  Sech-shaped inter-particle potentials
and confined in slightly elongated two- and three-dimensional harmonic traps.

Summarizing, we predict that the density of trapped repulsive ultra-cold bosons
inevitably fragments into multi-hump structures to minimize strong inter-particle repulsion.
The physics behind is an interplay between classical ``electrostatic" repulsion 
which pushes the bosons from the trap center towards its edges, provoking thereby the formation of multi-hump structures in the density,
and quantum mechanics which governs loss of inter-hump coherence, i.e., fragmentation.

Finally, for experimental verification of the predicted phenomena with ultra-cold systems one needs that
(i) the range of inter-particle interaction
should be comparable with the length of trapping potentials;
(ii) repulsive interaction should be strong enough.
Both these requirements are already reachable within presently available technologies
in dipolar ultra-cold atomic/molecular BECs  trapped in tight optical traps and 
in trapped ultra-cold systems of ``dressed" Rydberg atoms, see \cite{SM} for discussion.

\subsection*{Acknowledgment}
I am grateful to Oksana I. Streltsova for 2D and 3D MCTDHB simulations, to Ofir E. Alon, Lorenz S. Cederbaum, Luis Santos
and Michael Fleischhauer for fruitful discussions.
Computation time on the bwGRiD and K100 clusters and financial support
within the framework of the ``Enable fund'' of the excellence initiative at Heidelberg university are greatly acknowledged.

\newpage
\thispagestyle{empty}

%
%
%
%
%
%
%
%


\makeatletter
\addtocounter{figure}{-4}
\renewcommand{\figurename}{Figure S\hglue -0.12 truecm}
\renewcommand*{\@biblabel}[1]{[S#1]}
\makeatother

\section*{\large Supplementary Material} 

In this supplementary material we show how the Multi-Configurational Time-Dependent Hartree for Bosons
(MCTDHB) theory attacks time-dependent and static problems, and
how the many-body theory boils down to the multi-orbital mean-field.
These relations allow us to understand that 
the self-induced effective potentials $V^{eff}_{\phi_i}(\r)$
introduced in the main manuscript at the mean-field level have the many-body origin behind.
Then we show how to analyze the many-body solutions at hand and
demonstrate how a normal (standard) condensate is transforming to  multi-hump
fragmented state with an increase of the inter-particle repulsion.
Finally, we address the question  whether it is possible to verify the predicted physics within 
presently available experimental setups.

\section*{MCTDHB equations-of-motion}

The Multi-Configurational Time-Dependent Hartree method for Bosons (MCTDHB)
is used to solve the time-dependent Schr\"odinger equation:
$$
 \hat H \Psi = i \frac{\partial \Psi}{\partial t}.
$$
Here $\hbar=1$ and $\hat H$ is the generic many-body Hamiltonian defined in the main text.
The  MCTDHB {\it ansatz} for the many-body wave-function $\Psi(t)$ is taken as a linear combination of time-dependent permanents:
$$
\left|\Psi(t)\right> =
\sum_{\vec{n}}C_{\vec{n}}(t)\left|\vec{n};t\right>,   \eqno{\mathrm {(S0)}}
$$
where the summation runs over all possible configurations whose
occupations $\vec{n}= (n_1,n_2,n_3,\cdots,n_M)$ preserve the total number of bosons $N$.
So, the Fock state $|n_1,n_2\rangle$ studied in the main text is one of the terms in this sum.
The expansion coefficients $\{C_{\vec{n}}(t)\}$ and shapes of the orbitals $\{\phi_k(\br,t)\}$ 
are variational time-dependent parameters of the MCTDHB method,
determined by the time-dependent variational principle.

To solve a time-dependent problem means to specify an initial many-body state and to find how the many-body function evolves.
The initial MCTDHB many-body state is given by initial expansion coefficients $\{C_{\vec{n}}(t=0)\}$ and by initial shapes of the orbitals $\{\phi_k(\r,t=0)\}$.
To determine their evolution one has to solve the respective governing equations which we list below.
The equations-of-motion for expansion coefficients $\{C_{\vec{n}}(t)\}$ read:
$$
 {\mathbf H}(t)\C(t) = i\frac{\partial \C(t)}{\partial t}, \eqno{\mathrm {(S1)}}
$$
where ${\mathbf H}(t)$ is the Hamiltonian matrix  with elements
$H_{\vec{n}\vec{n}'}(t) = \left<\vec{n};t\left|\hat H\right|\vec{n}';t\right>$.

The equations-of-motion for the orbitals $\phi_j(\r,t)$, $j=1,\ldots,M$ are:
$$
  i\left|\dot\phi_j\right> = \hat {\mathbf P} \left[\hat h \left|\phi_j\right>  +\lambda_0 \sum^M_{k,s,q,l=1}
  \left\{\brho(t)\right\}^{-1}_{jk} \rho_{ksql} \hat{W}_{sl} \left|\phi_q\right> \right], \eqno{\mathrm {(S2)}} 
$$
where  $\hat {\mathbf P} = 1-\sum_{j'=1}^{M}\left|\phi_{j'}\left>\right<\phi_{j'}\right|$ is the projector,
$\rho_{kq}$ and $\rho_{ksql}$ are elements of the reduced
one- and two-body density matrices available at every point in time from
the many-body wave-function $\Psi(t)$:
$$
 \rho(\r_1|\r'_1;t) =  N\int \Psi^\ast(\r'_1,\r_2,\ldots,\r_N;t)
 \Psi(\r_1,\r_2,\ldots,\r_N;t) d\r_2 d\r_3 \cdots d\r_N = \nonumber
 $$
 $$ 
= \left<\Psi(t)\left|\hat{\mathbf \Psi}^\dag(\r'_1)\hat{\mathbf \Psi}(\r_1)\right|\Psi(t)\right> =
 \sum^M_{k,q=1} \rho_{kq}(t) \phi^\ast_k(\r'_1,t)\phi_q(\r_1,t),  \eqno{\mathrm {(S3)}}
$$
$$
\!\!\! \rho(\r_1,\r_2|\r'_1,\r'_2;t) =
N(N-1)\int\Psi^\ast(\r'_1,\r'_2,\r_3,\ldots,\r_N;t) \Psi(\r_1,\r_2,\r_3,\ldots,\r_N;t)
     d\r_3 \cdots d\r_N = \nonumber 
$$
$$ 
\!\!\! =\left<\Psi(t)\left|\hat{\mathbf \Psi}^\dag(\r'_1)\hat{\mathbf \Psi}^\dag(\r'_2)
\hat{\mathbf \Psi}(\r_1)\hat{\mathbf \Psi}(\r_2)\right|\Psi(t)\right> =
 \sum^M_{k,s,q,l=1} \rho_{ksql}(t)
\phi^\ast_k(\r'_1,t) \phi^\ast_s(\r'_2,t) \phi_q(\r_1,t) \phi_l(\r_2,t). \nonumber 
$$
and 
$$
  \hat W_{sl}(\r,t) = \int \phi_s^\ast(\r',t) W(\r-\r') \phi_l(\r',t) d\r' \eqno{\mathrm {(S4)}}
$$
are {\it local} time-dependent potentials which we analyze later.

\subsection*{From time-dependent MCTDHB to static MCHB}

Here we are interested in static solutions, i.e., in many-body eigenstates of the time-independent Schr\"odinger equation.
Within the MCTDHB theory they are obtained by propagating the MCTDHB equations-of-motion in imaginary time.
This procedure relaxes an initial state to an eigenstate, typically to the ground-state, and, thus, is called relaxation.
To find many-body eigenstates of the time-independent Schr\"odinger equation, i.e., static solutions,
one can boil down the full time-dependent MCTDHB theory to its time-independent version --
to Multi-Configurational Hartree for Bosons (MCHB) method [27].
From mathematical point of view this is done by switching off all time-dependency 
in the respective equations of motion (S1, S2).

The equations-of-motion for the expansion coefficients Eq.~(S1) are reduced to standard diagonalization procedure:
$$
 {\mathbf H}\C = E \C, \eqno{\mathrm {(S5)}}
$$
where $E$ is the total energy corresponding to the eigenstate $\C$.
The static version of the equations-of-motion (S2) can be rewritten without projections:
$$
\hat h \left|\phi_j\right>  +  \lambda_0 \sum^M_{k,s,q,l=1} \left\{\brho\right\}^{-1}_{jk} \rho_{ksql} \hat{W}_{sl} \left|\phi_q\right>  
= \sum^M_{i=1} \mu_{ji} \left|\phi_i\right>,   \eqno{\mathrm {(S6)}}
$$
where $\mu_{ji}$ are scaled Lagrange multipliers introduced to ensure orthonormality of the orbitals (fragments) $\phi_j(\r)$.
The optimal shapes of the now time-independent orbitals $\phi_j(\r)$, $j=1,\ldots,M$ and optimal
$\C$ have to be determined by solving the coupled system of Eqs.~(S5, S6) self-consistently. 

\subsection*{From many-body to mean-field}
Now our goal is to demonstrate how the full many-body MCTDHB  theory boils down to the multi-orbital (Best) Mean-Field (BMF).
From formal mathematical point of view this reduction
is reached by taking a single term (permanent) in the MCTDHB ansatz  Eq.~(S0) and putting all other to zero.
Physically it means that the many-body solution is assumed to be well-described by $M$ separated sub-clouds (fragments)
carrying $n_1,n_2,n_3,\ldots,n_M$ bosons. 
Moreover, it also implies that there is no exchange of bosons between  the fragments. 

The multi-orbital mean field ansatz reads:
$$
\left|\Psi\right> = \left|n_1,n_2,\cdots,n_M\right>.
$$
The MC(TD)HB equations-of-motions for the expansion coefficients Eqs.~(S1, S5) 
become redundant, because we have only a single expansion coefficient.
All off-diagonal elements of $\rho_{kq}$ and $\rho_{ksql}$ corresponding to the single-permanent state are zero, while
the diagonal ones are known in advance  $\rho_{ii}=n_i$,  $\rho_{iiii}=n_i^2-n_i$ and $\rho_{ijij}=\rho_{ijji}=n_in_j$.
The above facts considerably simplify the static equations-of-motion (S6) for orbitals $\phi_j(\r)$, $j=1,\ldots,M$:
$$
\left[ \hat h  + (n_j-1) \lambda_0 \hat{W}_{jj} + \sum^M_{i \ne j} n_i  \lambda_0 \hat{W}_{ii} \right] \left|\phi_j\right>  
+\sum^M_{i \ne j} n_j  \lambda_0 \hat{W}_{ij} \left|\phi_i\right>  
= \sum^M_{i=1} \mu_{ji} \left|\phi_i\right>.    \eqno{\mathrm {(S7)}}
$$
Hence, for a given set of the occupations numbers $(n_1,n_2,\cdots,n_M)$ by solving this system one determines the optimal shapes of the fragments.

\subsection*{Role of  $\hat{W}_{sl}(\r,t)$}
The {\it local} potentials $\hat{W}_{sl}(\r,t)$ defined in Eq.~(S4) are generally  complicated time-dependent objects
which depend on inter-particle interaction and on shapes of the involved time-dependent orbitals $\phi_j(\r,t)$, $j=1,\ldots,M$.
However, in the static cases, see Eqs.~(S6, S7), they become time-independent functions:
$$
  \hat W_{sl}(\r) = \int \phi_s^\ast(\r') W(\r-\r') \phi_l(\r') d\r'.   \eqno{\mathrm {(S8)}}
$$
In the present study the mean-field physics dictates localization of the fragments $\phi_j(\r)$, $j=1,\ldots,M$ 
at different places of the trap potential (see e.g. Fig.~2). It also means that the overlap between the different fragments is negligible.
Hence, contribution from $\hat W_{sl}(\r)$ terms in $s\ne l$ cases can be neglected,
because the integral over two functions with zero overlap is zero. 
This fact helps us to derive from Eqs.~(S7) the mean-field equations for a two-fold fragmented system
with well-isolated left and right fragments discussed in the main text.
To facilitate a formal connection between external trapping potential $V(\r)$  and
effective potentials we redefine them as $V^{eff}_{\phi_i}(\r) \equiv \hat W_{ii}(\r)$, with $i=L,R$.

The Multi-Configurational Time-Dependent Hartree method has been successfully formulated [S1]
in terms of reduced one- and two-body densities and {\it local} potentials $\hat{W}_{sl}(\r,t)$
also for systems of mixed bosons,  fermions as well as for mixture thereof. 
So, the proposed analysis of the multi-hump solutions in terms of 
{\it local} self-induced effective potentials $V^{eff}_{\phi_i}(\r)$
is expected to be universal and can be applied to all many-body systems.

\section*{Natural analysis of the many-body wave-functions}

Having the MC(TD)HB($M$) many-body wave function $\left|\Psi\right>$ at hand
we can analyze it.
The reduced one-body density matrix available from Eq.~(S3) can be diagonalized:
\beq
\rho(\r,\r';t)=\sum^M_{k,q=1} \rho_{kq}(t)\phi^\ast_k(\r',t)\phi_q(\r,t) = \sum^M_{k=1} n_k(t) \phi_k^{\ast NO}(\r',t) \phi_k^{NO}(\r,t).
\nonumber
\eeq
The obtained eigenvalues $n_k$ and eigenvectors $\phi_k^{NO}$ are called
natural occupation numbers and natural orbitals, respectively.
This diagonalization procedure is often referred to as natural analysis.
The natural occupation numbers $n_k$ can be considered as average numbers of bosons residing in $\phi^{NO}_k$.
The natural orbital analysis is used to characterize the system:
the system is condensed [S2] when only a single natural orbital
has a macroscopic occupation and {\it fragmented} if several natural orbitals are macroscopically occupied  [10].
Usually, one plots only the diagonal part $\rho(\r;t)\equiv\rho(\r,\r;t)$ often referred to as the density.

\begin{figure}[]
\includegraphics[width=17cm,angle=0]{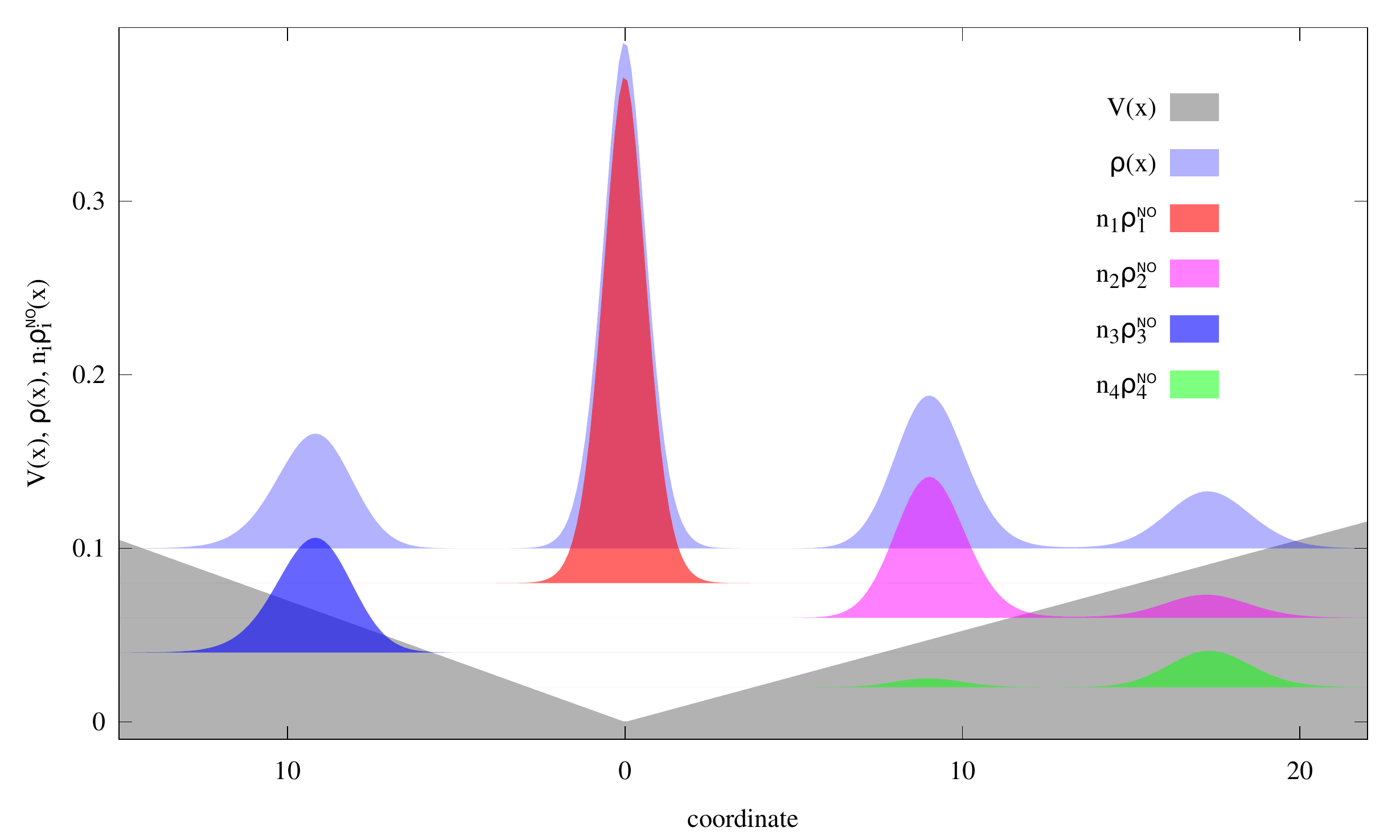}
\caption{{\bf Decomposition of the reduced one-body density $\rho(x)$ into natural fragments.}
The four-hump density of $N=108$ bosons interacting via $W(\bR)=1/\sqrt{(\frac{|x-x'|}{D})^{4}+1})$ of range $D=5$ with $\lambda_0=0.3$
in asymmetric linear trap is formed by four contributing natural sub-fragments $\rho(x)=\sum^4_{i=1}n_i \rho^{NO}_i(x)$, 
$|\phi^{NO}_i|^2\equiv \rho^{NO}_i$.
The first fragment, carrying $n_1=52$ bosons is localized at the trap minimum and forms the most intense hump.
The left-most hump is formed by $n_3=20$ bosons residing in the third well-localized natural orbital.
The two other humps at the right are formed by $n_2\approx27.6$ and $n_4\approx8.4$ bosons
residing in the second and fourth natural orbitals which are slightly delocalized.  
Shown are normalized reduced one-particle and partial densities corresponding to the natural fragments which are shifted vertically for a better visualization.
The trap potential is also rescaled to fit the figure.
All quantities shown are dimensionless.}
    \label{fig1.sup}
\end{figure}

In Fig.~S1 we plot the results of the natural analysis done for the ground state (static solution) 
of the system with $N=108$ bosons 
interacting via $W(\bR)=1/\sqrt{(\frac{|x-x'|}{D})^{4}+1})$ of range $D=5$ and $\lambda_0=0.3$
confined in asymmetric linear trap $V(x)=\{-x: x<0; \, 3x/4: x\ge0 \}$, studied in the main text.
The ground-state density obtained from MCTDHB(4) solution has four well-pronounced picks/humps. 
The natural analysis applied shows that this density is formed by four contributing natural sub-fragments,
i.e., $\rho(\r)=\sum^4_{i=1}n_i |\phi^{NO}_i(\r)|^2$.
In Fig.~S1 the normalized densities of the natural fragments, $n_i/N \rho^{NO}_i(\r)$ ($|\phi^{NO}_i|^2\equiv \rho^{NO}_i$)
are shifted vertically for better visualization. The trap potential is also rescaled to fit the figure.
The natural orbitals are sorted according to the numerical values of the computed natural occupation numbers.
The first fragment with $n_1=52$ bosons is localized at the trap minimum and forms the most intense hump.
The left-most hump is formed by $n_3=20$ bosons residing in the third well-localized natural orbital.
Two other humps at the right are formed by $n_2\approx27.6$ and $n_4\approx8.4$ bosons
residing in the second and fourth natural orbitals. The two left fragments are well separated from the right ones.
The non-integer occupations of these right fragments indicate that the physics here is slightly beyond mean-field.
Indeed, the sub-clouds forming these humps are not completely localized, i.e., the respective natural orbitals have two-hump profiles
and non-zero density between the humps. This means that there is an exchange/flow of the bosons between the fragments.

Summarizing, in this example the many-body ground state is four-fold fragmented. 
The phase information between two left fragments is completely lost because they are very-well separated.
The two right-most fragments are not entirely isolated from each other and, therefore,  can communicate.
The phase contrast experiments where the quantum state is repeatedly prepared and released to form the interference patterns
would reveal different phase distribution patterns for the left and right sub-clouds.
A perfect random phase distribution between the left fragments would contrast with
a deformed phase distribution between the right sub-clouds.

\section*{From a normal condensate to multi-fold fragmented states}

Let us now show how an increase of the strength $\lambda_0$ of the inter-particle 
interaction modifies the ground state properties of the trapped systems. 
In Fig.~S2 we plot the results of the natural analysis done for the ground state (static solution) 
of the same system as in Fig.~S1 for several other values of $\lambda_0=0.01,0.1,0.2$.
For weak interactions the normal condensation is of course recovered
i.e., only single natural orbital is macroscopically occupied $n_1\approx 100\%$.
However, for stronger repulsive finite-range interactions
to minimize the total energy the density profile develops multi-hump structures and the systems become multi-fold fragmented.
For $\lambda_0=0.1$ two-fold (two-hump) fragmented state is formed by two
localized fragments with $n_1\approx91.1\%$ (left) and $n_2\approx8.9\%$ (right) bosons correspondingly.
For $\lambda_0=0.2$ three localized fragments with $n_1\approx61.2\%$ (center),  $n_2\approx24.9\%$  (right) and $n_3\approx13.9\%$ (left)
form three-fold (three-hump) fragmented state.

\begin{figure}[]
\includegraphics[width=17cm,angle=0]{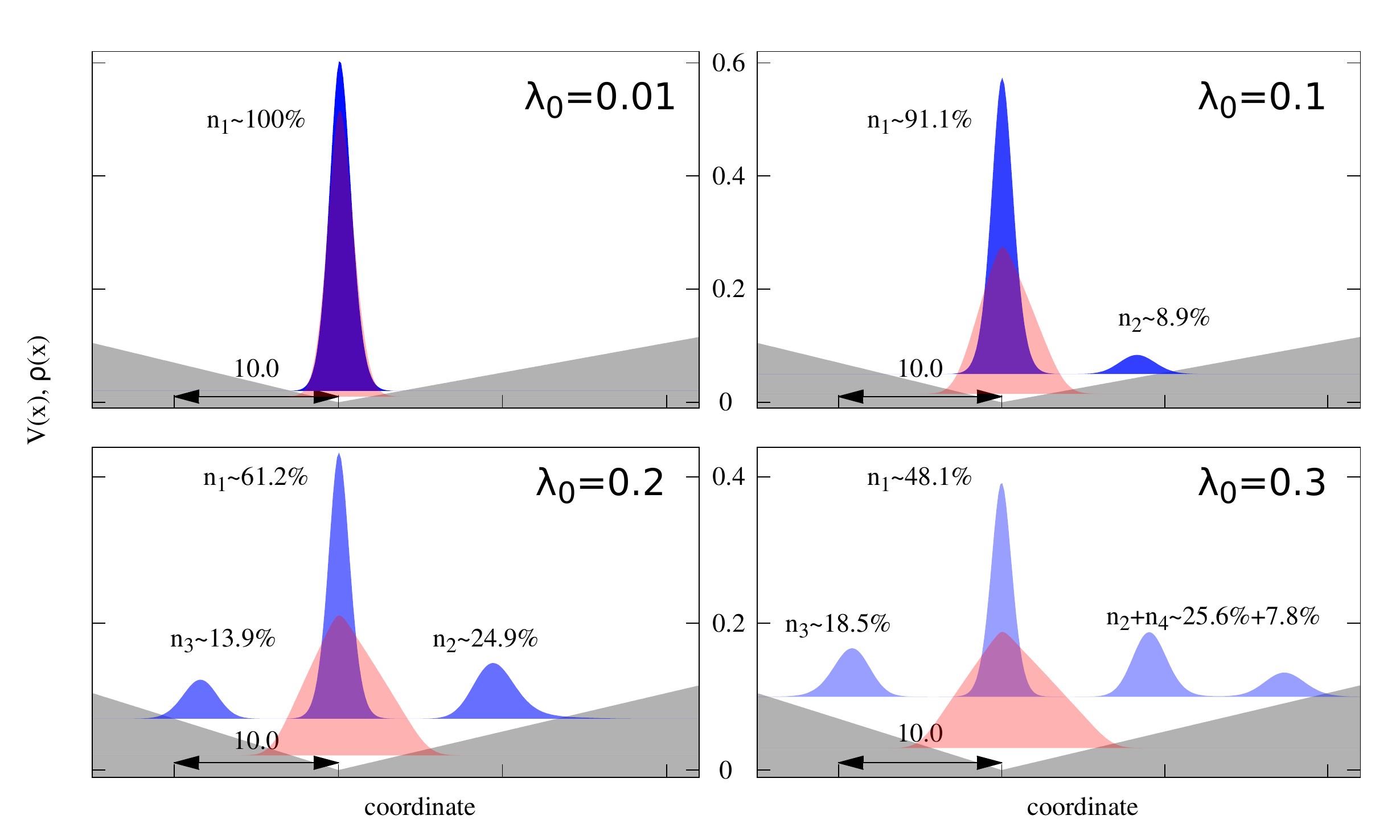}
\caption{{\bf From a normal condensate to multi-fold fragmented states.}
The densities $\rho(x)$ of $N=108$ bosons confined in the asymmetric linear trap (same as in Fig.~S1) 
are computed for several different interaction strengths $\lambda_0=0.01,0.1,0.2,0.3$.
The results obtained for $W(\bR)=1/\sqrt{(\frac{|x-x'|}{D})^{4}+1})$ interaction potential of range $D=5$
and a pure contact interaction  $W(\bR)=\delta{|x-x'|}$ are contrasted.
In the regime of a weak repulsion $\lambda_0=0.01$ irrespective to the particular form of interactions
both systems have a single-hump density profile and are fully condensed $n_1\approx 100\%$. 
For a stronger repulsion $\lambda_0=0.1$ the contact interaction gives a condensed single-hump state
while the finite-range interaction -- a two-fold (two-hump) fragmented state with 
localized fragments occupied by $n_1\approx91.1\%$ and $n_2\approx8.9\%$ bosons.
For stronger repulsion $\lambda_0=0.2$ to minimize the total energy the system with finite-range interaction
forms a three-hump (three-fold) fragmented state,  the contact interaction still result in a single-hump condensate.
The structure of the four-fold fragmented ground state for the finite-range interaction with $\lambda_0=0.3$ is discussed in more detail in Fig.~S1.
The corresponding system with contact interaction is a slightly depleted single-hump condensate $n_1\approx 98.9\%$.
The normalized reduced one-particle densities $\rho(x)$ are shifted vertically for a better visualization.
The trap potential $V(x)$ is also rescaled to fit the figure. 
All quantities shown are dimensionless.}
    \label{fig2.sup}
\end{figure}

For comparison we compute the ground state for the same systems but with a pure contact inter-particle interaction  $W(\bR)=\delta{(x-x')}$.
In all the presented cases, i.e.,  $\lambda_0=0.01,0.1,0.2,0.3$ the systems with contact interaction 
are essentially condensed and have single-hump densities. 
The condensed fraction of the system with N=108 bosons and  $\lambda_0=0.3$ is  $n_1\approx 98.9\%$.
For smaller $\lambda_0$ the condensed fractions are even larger.

\section*{On possible experimental realization of the predicted phenomena}

To observe the predicted effects the first necessary condition is that the length/range of the inter-particle interaction
should be comparable with the length of trapping potentials. 
The second necessary condition is that the interaction should be strong enough.
There are at least two generically different ultra-cold systems where one expects to observe 
the formation of few-hump fragmented ground states: the dipolar condensates 
made of atomic or molecular species with non-zero electric/magnetic moments 
and ultra-cold systems made of Rydberg-dressed atoms [10, 11, S3] 

Let us first consider the dipolar condensates.  
The dipolar systems apart from the long-range dipole-dipole interaction (DDI) also posses
a residual contact contribution, so, in this respect, the main obstacle in the preparation of a strongly interacting dipolar system 
is the contact term. However, luckily, the short- and long-range terms do originate to different physics and, hence,
can be individually manipulated. It was proposed to use the famous Feshbach-resonance technique 
to demolish the contribution of the short-range terms, providing
thereby opportunity to study pure dipolar systems. The interaction between two polarized (oriented) dipoles reads:
$$
\tilde{W}_{dd}(|\tilde{r}_i-\tilde{r}_j|)=\frac{d^2}{\tilde{r}_{ij}^3} (1-3 \cos^2 \theta ),   \eqno{\mathrm {(S9)}}
$$
where $\tilde{r}_{ij}$ is the separation between the dipoles, $\theta$ defines their orientation
and $d^2$ is the strength of the dipole-dipole interaction.
$d^2=\frac{\tilde{D}^2}{4 \pi \epsilon_0}$ for electric dipoles and
$d^2=\frac{\mu_0 \mu^2}{4 \pi}$ for magnetic ones. Here
$\tilde{D}$ -- is electric dipole moment in Debye, $\mu=g_L \mu_B$,
$g_L$ is the Land\'e factor, $\mu_0= 4 \pi \times 10^{-7} V·s/(A·m)$ is the vacuum permeability,
$\epsilon_0 = 8.854187817 \times 10^{-12} F m^{-1} $ is the dielectric permittivity of classical vacuum,
$\mu_B=9.27400968 \times 10^{-24} JT^{-1}$ is the Bohr magneton.  

To connect these ``real" inter-particle interaction potentials Eq.~(S9)
to ``dimensionless" ones used in the main text we introduce 
a change of the variables $\tilde{r}_{i} =r_i l_{\omega}$.
Here $l_{\omega}=\sqrt{\frac{\hbar}{m \omega}}$ is the length of the external harmonic trap of frequency $\omega$
associated with the size of the harmonic oscillator ground-state wave function, $m$ -- the mass of a trapped particle.
This change of variables rescales the Hamiltonian $\hat{H}/ (\hbar \omega)$ and respective  kinetic energy
$-\frac{\hbar}{2 m} \nabla^2_{\tilde{r}_i} \to -\frac{1}{2}\nabla^2_{r_i}$, 
single-particle trapping potential $\frac{m \omega}{2} \tilde{r}^2_i \to \frac{1}{2} r^2_i$, 
and, the inter-particle interaction
$$
W_{dd}(|r_i-r_j|)=\frac{\lambda_0}{r_{ij}^3} (1-3 \cos^2 \theta).  \eqno{\mathrm {(S10)}}
$$
Here $\lambda_0=d^2/ (\hbar \omega l^3_{\omega})$ is the rescaled interaction strength [S4]. 

The length scale (range) associated to the pure DDI is $l_{dd}=d^2 m /\hbar$ [S4]. 
Taking the values of magnetic moments $\mu=6 \mu_B$ for $^{52}Cr$ and $\mu=7 \mu_B$ for $^{164}Dy$  and $^{168}Er$  
one obtains the respective DDI length scales (ranges): 
$l^{Cr}_{dd} \sim$ \SI{0.0024}{\micro\metre},
$l^{Dy}_{dd} \sim$ \SI{0.0103}{\micro\metre},
$l^{Er}_{dd} \sim$ \SI{0.0106}{\micro\metre}.
For these atoms a typical trap with  $\omega=2\pi \times$\SI{100}{\hertz} would result in $l_{\omega} \sim$ 0.5-\SI{1.5}{\micro\metre}, i.e.,
the length of this trap is few orders of magnitude larger then the $l_{dd}$.
Fortunately, with optical lattices one can create traps with much stronger confinement.
For example, in recent experiments with Cr reported in Ref.~[S5], 
the vibrational spacing in a lattice was about \SI{130}{\kilo\hertz}.
This lattice confinement results in $l_{\omega_L}\sim$ \SI{0.035}{\micro\metre} of the corresponding
trap length and much better ratio $ l_{\omega_L} \sim 16 l^{Cr}_{dd}$.
If a similar confinement could be created for Dy and Er atomic clouds, one would get:
$l_{\omega_L} \sim 2.8 l^{Dy}_{dd}$ and $l_{\omega_L} \sim 2.77 l^{Er}_{dd}$ correspondingly.
Hence,  the necessary ratio between the lengths of the DDI and of the confinement potentials
can be reached in modern experiments with dipolar Dy and Er BECs in optical lattices.
Let us now verify how strong are the DDIs in these systems.
The \SI{130}{\kilo\hertz} lattice-induced trap
rescales the inter-particle DDI interaction strengths $\lambda_0=d^2/ (\hbar \omega l^3_{\omega})$ and results in:
$\lambda_0 \sim 0.062$ for Cr,  $\lambda_0 \sim 0.474$ for Dy and  $\lambda_0 \sim 0.491$ for Er. It is worthwhile to recall that the
multi-fold fragmented states depicted in Figs.~S1,S2 were reached with $\lambda_0\sim0.1-0.3$.

The electric DDI can, in principle, be realized with ultra-cold clouds (BEC) of oriented molecules.
According to Ref.~[S4] 
the DDI length of polar molecules of mass 100 a.m.u 
with an electric dipole of 0.2 Debye  is $l^{Mol}_{dd} \sim$ \SI{0.03}{\micro\metre}.
A lattice-induced  \SI{10}{\kilo\hertz} trap would result in $l_{\omega_L} \sim$ \SI{0.1}{\micro\metre}, so,
formally, $l_{\omega_L} \sim 3 l^{Mol}_{dd}$. The strength of the electric DDI in this trap is $\lambda_0 \sim 0.2$.
A tighter trap would results in a shorter $l_{\omega_L}$ and stronger DDI.

Concluding, the predicted multi-fold fragmented states can, in principle, be observed within presently available technologies
in dipolar ultra-cold atomic/molecular BECs  trapped in tight optical traps.

Let us now consider the opportunity to realize the predicted multi-hump muti-fold fragmented physics in Rydberg-dressed systems.
The off-resonant optical coupling of ground state atoms to highly excited Rydberg states [10, 11] 
allows to modify and control the shape and strength of effective two-body interactions.
The advantage of this technique is that a small component of the Rydberg state is admixed to the ground state of atoms,
providing thereby an additional degree of manipulation of the interaction strength.
The shape of the potential energy of the two many-electron atoms excited to a Rydberg state depends very much on details of the electronic structure.
The most common long-range behavior, however, is of a standard Van der Waals ($C_6/R^6$) type with possible admixture of terms of other degrees, 
e.g., a pure dipole interaction  ($C_3/R^3$). These additional contributions are responsible for the tails of the interaction potentials.
The ``dressing" of two Rydberg atoms with van der Waals type of interaction results
Refs.~[10, 11]  
in a two-body effective inter-particle potential:
$$
\tilde{W}_{dd}(|\tilde{r}_i-\tilde{r}_j|)=\frac{\tilde{C}_6}{\tilde{r}_{ij}^6+R_c^6}=\frac{\Omega^4}{8 \Delta^3} \frac{\hbar}{(\tilde{r}_{ij}/R_c)^6+1}.
 \eqno{\mathrm {(S11)}}
$$
A change of variables $\tilde{r}_{i} = r_i l_{\omega}$
allows us to rewrite it in the units of external confining potential $l_{\omega}=\sqrt{\frac{\hbar}{m \omega}}$ as:
$$W(|r_i-r_j|)=\frac{\lambda_0}{(r_{ij}/D)^6+1},  \eqno{\mathrm {(S12)}}$$ 
where $\lambda_0=\frac{\hbar\Omega^4}{(2 \Delta)^3}\frac{1}{\hbar \omega}$ is the rescaled interaction strength
and $D=R_c$. In the present study we have used a similar 
one-dimensional inter-particle interaction $\lambda_0/\sqrt{(\frac{|x-x'|}{D})^{4}+1})$, 
see Figs.~1a,1d and Figs.~S1,S2.
The strength $\lambda_0$ of the inter-particle interaction depends on 
experimentally tunable parameters: $\Omega$  -- a two-photon Rabi frequency between the involved atomic levels,
$\Delta$ -- detunings of lasers with respect to the atomic transitions, and the external confinement $\omega$.
The ``screening" constant $D\equiv R_c=\frac{C_6}{2 \hbar |\Delta|}$
defines the critical distance, below which the interaction is constant, i.e., originates to the blockade phenomenon [S3].
This range depends on the detuning $\Delta$ and on 
a pure spectroscopic $C_6$ coefficient governed by the electronic structure of the excited state, i.e., it
can be manipulated by a proper choice of an atomic Rydberg level.

What are the presently available/reachable experimental conditions for Rydberg excitations? 
In recent experiments Ref.~[S6]  with Rb BECs 
a few tens of Rydberg atoms in a quasi-1D trap have been successfully detected.
The reported blockade radii were between 5-\SI{15}{\micro\metre} 
while the radial dipole trap frequencies used were around \SI{100}{\hertz} resulting in radial dimension of
order of 1-\SI{2}{\micro\metre}, i.e., ratio $l_{\omega} \sim 0.1 R^{Rb}_{c}$.
It means that the range of the interaction was larger then the the size of the trap in the transverse direction. 
The strength of the interaction of the dressed Rydberg atoms
is defined by $\frac{\Omega^4}{8 \Delta^3}$ ratio and can be tuned between 0.1-\SI{10}{\kilo\hertz}, implying that 
in a \SI{100}{\hertz} trap the corresponding dimensionless interaction strength $\lambda_0=\frac{\hbar\Omega^4}{(2 \Delta)^3}\frac{1}{\hbar \omega}$
can be $\sim$ 1-100. 
So, formally, the lengths/ranges and interaction strength needed to observe the predicted multi-hump muti-fold fragmented states 
are already reachable with ultra-cold ``dressed" Rydberg atoms.

Finally, the results discussed in this work have been nicely explained at the multi-orbital mean-field level in terms of self-induced effective barriers.
This also implies that the underlying physics is defined not by a strong interaction strength $\lambda_0$ between individual particles 
but, rather, by the $\lambda_0 N$ interaction mean-field parameter (also known as non-linearity). 
So, the interesting physics of the multi-hump fragmented states reported and presented in this study for $N=108$ bosons
can, in principle,  be also realized in the systems with larger particle numbers and weaker inter-particle  interactions.

\end{document}